\documentclass[a4paper,11pt]{article}
\usepackage{pos}
\usepackage{floatflt}

\usepackage{etoolbox}
\patchcmd{\thebibliography}
  {\settowidth}
  {\setlength{\itemsep}{0pt plus 0.1pt}\settowidth}
  {}{}
\apptocmd{\thebibliography}
  {\small}
  {}{}

\title{Gamma-ray signatures from pair cascades in recombination-line radiation fields}
 \ShortTitle{Gamma-ray signatures from pair cascades}

\author*[a]{\href{https://ui.adsabs.harvard.edu/public-libraries/kndW8kJdS7G7Pjf7W_2DAA}{Christoph Wendel}}
\author[b]{\href{https://orcid.org/0000-0002-6729-9022}{Josefa Becerra Gonz\'{a}lez}}
\author[c]{\href{https://orcid.org/0000-0002-5656-2657}{Amit Shukla}}
\author[d]{\href{https://www.mpp.mpg.de/david-paneque/}{David Paneque}}
\author[a]{\href{https://www.physik.uni-wuerzburg.de/astro/mitarbeiter/ag-mannheim/}{Karl Mannheim}}

\affiliation[a]{Julius-Maximilians-Universit\"at W\"urzburg, Germany}\emailAdd{cwendel@astro.uni-wuerzburg.de}

\affiliation[b]{Instituto de Astrof\'isica de Canarias and Universidad de La Laguna, Spain}

\affiliation[c]{Indian Institute of Technology Indore, India}

\affiliation[d]{Max-Planck-Institut f\"ur Physik, Germany\\}

\emailAdd{jbecerra@iac.es}
\emailAdd{amit.shukla@iiti.ac.in}
\emailAdd{dpaneque@mpp.mpg.de}
\emailAdd{mannheim@astro.uni-wuerzburg.de}

\abstract{Beams of ultra-relativistic electrons in blazar jets develop pair cascades interacting with ambient soft photons. Employing coupled kinetic equations with escape terms, we model the unsaturated pair cascade spectrum. We assume that the gamma rays predominantly scatter off recombination-line photons from clouds photoionised by the irradiation from the accretion disk and the jet. The cascade spectrum is rather insensitive to the injection of hard electron spectra associated with the short-time variability of blazars. Adopting physical parameters representative of Markarian 501 and 3C 279, respectively, we numerically obtain spectral energy distributions showing distinct features imprinted by the recombination-line photons. The hints for a peculiar feature at $\sim 3\,\rm TeV$ in the spectrum of Markarian 501, detected with the MAGIC telescopes during a strong X-ray flux activity in 2014 July, can be explained in this scenario as a result of the up-scattering of line photons by beam electrons and the low pair-creation optical depth. Inspecting a high-fidelity {\it Fermi}-LAT spectrum of 3C 279 from January 2018 reveals troughs in the spectrum that coincide with the threshold energies for gamma rays producing pairs in collisions with recombination-line photons, and the absence of exponential attenuation. Our finding implies that the gamma rays in 3C 279 escape from the edge of the broad emission line region.}

\FullConference{37$^{\rm{th}}$ International Cosmic Ray Conference (ICRC 2021)\\
		July 12th -- 23rd, 2021\\
		Online -- Berlin, Germany}

\begin{document}

\maketitle

\section{Introduction}\label{SecIntroduction}

Active galactic nuclei (AGNs) with a relativistic jet pointing towards the observer are called blazars. 
Broadband spectral energy distributions (SEDs) of blazars extend over up to 20 orders of magnitude and show two wide bumps. The first bump is in the range $\approx 10^{12}\,\rm{Hz} - 10^{18}\,\rm{Hz}$ and is usually ascribed to synchrotron radiation of non-thermal electrons moving relativistically in a magnetised region, which itself propagates along the jet with relativistic velocity. The second bump is situated at MeV to TeV energies. In the framework of leptonic models, its cause is explained via the synchrotron self-Compton (SSC) model, meaning that the electrons not only emit synchrotron photons but also inverse-Compton (IC) up-scatter these same photons, or via the external Compton (EC) mechanism, meaning that the electrons IC up-scatter ambient photons from external sources \cite{2013LNP...873.....G,2017FrASS...4....6F}. In limited ranges of energy, blazar SEDs can often be described satisfactorily by simple functions like power laws, power laws with exponential cutoff or logparabolas. Additional SED substructure beyond these simple functions, as well as minute-scale variability, points to radiation processes beyond the SSC/EC scenario \cite{ReferenceVariability4,Ahnen18}.

In this contribution, we consider a cascade resulting from the interaction of non-thermal electrons or gamma rays with soft external photons from ionised clouds surrounding the jet base. Taking into account not only pure pair absorption but also reprocessing, the cascaded emission can provide explanations for a narrow SED feature in Markarian 501 (Mrk\,501
) as well as for an SED of 3C\,279, that is challenging to describe with usual functions
.

In the following, $m_{\rm e}$, $c$, $e$ and $\sigma_{\rm Th}$ denote the electron rest mass, the velocity of light, the elementary charge, and the Thomson cross section, respectively. Analogously to the Lorentz factor $\gamma$, the soft and gamma-ray photon energy, $x$ and $x_\gamma$, are measured in units of $m_{\rm e} c^2$.

\section{An electron beam causing a narrow feature at \textasciitilde 3 TeV in the gamma-ray emission of Mrk 501}\label{SecMrk501}

On MJD 56857.98 (2014 July 19), during a multi-wavelength campaign, hints of a peculiar spectral feature centred at $\sim 3\,\rm TeV$ (shown in Fig. \ref{FigMrk501SED}) were found in Mrk\,501 with the Major Atmospheric Gamma-Ray Imaging Cherenkov (MAGIC) telescopes \cite{2019ICRC...36..554B,2020A&A...637A..86M}. Power-law, logparabola and exponential power-law fits of this narrow bump can be discarded with a confidence level $\geq 3\,\sigma$. A likelihood ratio test shows that a two-component model with one broad and one narrow logparabola is preferred over a pure logparabola at $\sim 4\,\sigma$. A detailed assessment of the data analysis of this multi-wavelength observation and of the peculiar $\sim 3\,\rm TeV$ bump are presented in the Talk \#79 by Becerra Gonz\'{a}lez, J., of this conference. Besides evoking a two-zone SSC model with extreme parameters and a model with relativistic Maxwellian electron energy distribution due to electron pile-up in energy space, the following model can be applied to explain the possible narrow spectral component in the gamma-ray spectrum measured with MAGIC \cite{2020A&A...637A..86M}.

\subsection{Description of the setting}\label{SubsecSetting}

\begin{floatingfigure}[l]{0.51\textwidth}
\centering
\hspace{-1cm}\includegraphics[width=0.57\textwidth]{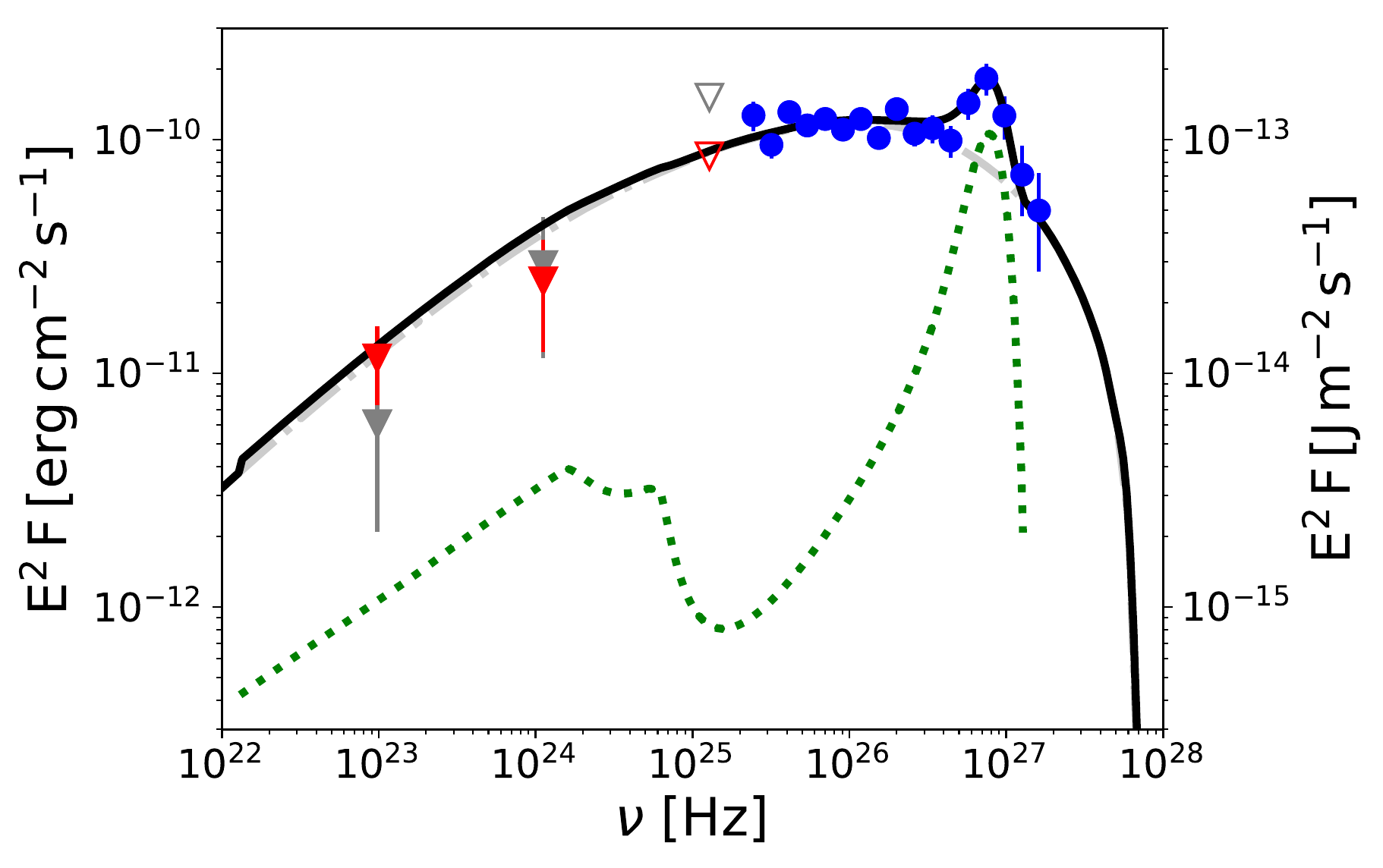}
\vspace{-0.3cm}\caption{SED of Mrk 501 from MJD 56857.98. The grey dot-dashed, the green dotted and the black solid line shows the SSC contribution, the cascaded flux contribution $F_{\rm{casc}}$ and the sum of both, respectively, multiplied with the squared photon energy $E$. The MAGIC spectral data points are the blue circles, while {\it Fermi}-LAT data are shown by red and grey triangles.}
\label{FigMrk501SED}
\end{floatingfigure}

According to our model sketched in Fig. \ref{FigSketch}, the narrow feature at $\sim 3\,\rm TeV$ can be a signature of vacuum gap activity in the magnetosphere of Mrk\,501 \cite{2017AIPC.1792e0026W,2020A&A...637A..86M,2021A&A...646A.115W}. Around a rotating black hole (BH) the co-rotating magnetic field induces an electric field. To keep the electric field force-free, the magnetosphere has to be filled with plasma, whose charge density has to be equal to the Goldreich-Julian charge density. If the true charge density differs from the Goldreich-Julian density, an electric field arises parallel to the magnetic field and the respective region is called a vacuum gap. Motivated by numerical simulations \cite[e.g.][]{HirotaniPu1,2020ApJ...902...80K,Chen2}, we assume that a vacuum gap is located at the magnetospheric poles, sporadically active and providing a potential drop $\Phi$.

In the gap, we assume a pair materialisation rate (number per unit time and per unit volume)
\begin{equation}
K_{\rm{PP}} = 0.2 \, \sigma_{\rm Th} \, n_{{\rm PP}}^2 \, c
\label{eqPP}
\end{equation}
due to Breit-Wheeler pair production resulting from collisions of gamma-ray photons of density $n_{{\rm PP}}(L_{\rm{ADAF},\,\nu}(\dot m, T_{\rm e}))$, originating from the surrounding advection-dominated accretion flow (ADAF), whose spectral luminosity $L_{\rm{ADAF},\,\nu}(\dot m, T_{\rm e})$ depends (after setting the black hole mass $M_{\rm BH} = 10^9 \, M_{\odot}$ and after employing reasonable inner and outer radial flow boundaries) only on the electron temperature $T_{\rm e}$, and on the dimensionless accretion rate $\dot m$ \cite{Mahadevan}. The materialising electrons (or positrons) are accelerated to ultra-relativistic energies with a maximum at $\approx e \, \Phi$, and propagate away from the central object perpendicularly to the equatorial plane. In or immediately behind the gap, the electrons interact repeatedly with photons from the ADAF. By this, they are multiplied through IC scattering, curvature radiation and pair production by the number $\mathcal{N}$. During this multiplication, the energy of one original electron is inherited to its child electrons. Hence, after they leave the domain of the ADAF photons, the multiplication of the electron beam ceases. The maximum electron energy is now $e \, \Phi / \mathcal{N}$ and the injection rate (number per unit time and per unit volume) is
\begin{equation}
K_{\rm{G}} = K_{\rm{PP}} \, \mathcal{N}.
\label{eqK}
\end{equation}
We parametrise the electron spectral injection rate $\dot N_{\rm i}(\gamma)$ after the in-gap and post-gap multiplication by a Gaussian function with mean Lorentz factor $\gamma_{\mathrm{mean}}$, width $\varsigma$, normalisation $K_{\rm{G}}$, lower cutoff $\gamma_{{\mathrm{i}},\,1} = \gamma_{\rm{mean}} - 3.0 \, \varsigma$ and upper cutoff
\begin{equation}
\gamma_{{\mathrm{i}},\,0} = \gamma_{\rm{mean}} + 3.0 \, \varsigma = e \, \Phi / \mathcal{N}.
\label{eqPhi}
\end{equation}

Besides driving pair production in the magnetosphere, the ADAF illuminates and ionises gas clouds surrounding the central object. We assume that a fraction $\xi$ of the total ADAF luminosity $L_{\rm{ADAF,\,tot}}(\dot m, T_{\rm e})$ is reprocessed by $N_{\rm{cl}}$ clouds of radius $R_{\rm{cl}}$ into emission line photons with total luminosity
\begin{equation}
L_{\rm{lines,\,tot}} = \xi \, L_{\rm{ADAF,\,tot}}(\dot m, T_{\rm e}).
\label{eqLlinestot1}
\end{equation}
$L_{\rm{lines,\,tot}}$ is the emission line luminosity of the ensemble of clouds in the AGN, and the concomitant spectral number density of soft photon near each cloud is $n_0(x)$. Into $n_0$ we include the He II Ly $\alpha$ line, as well as the H Ly series, Ly $\beta$ and Ly $\alpha$ line, the latter of which has been observationally confirmed with a luminosity of $L_{\rm{Ly}\,\alpha,\,\rm{obs}} = 5.2 \cdot 10^{33} \, \rm{W}$ \cite{Stocke}. We use line strengths based on observational and synthetic spectra \cite{2005MNRAS.361..919P,2017MNRAS.464..152A}. Furthermore, we use $N_{\rm{cl}} = 10$, reflecting that Mrk\,501 is an evolved AGN with lack of gas supplies from the galaxy centre.

From the total energy density $u_{\rm{lines,\,tot}}$ of line photons in the clouds, we have another constraint for the line luminosity:
\begin{equation}
L_{\rm{lines,\,tot}} = N_{\rm{cl}} \, 4 \pi R_{\rm cl}^2 \, c \, u_{\rm{lines,\,tot}}
\label{eqLlinestot2}
\end{equation}
Similarly, for the H Ly $\alpha$ luminosity we have
\begin{equation}
N_{\rm{cl}} \, 4 \pi R_{\rm cl}^2 \, c \, u_{\rm{Ly}\,\alpha} = L_{\rm{Ly}\,\alpha,\,\rm{theo}} = L_{\rm{Ly}\,\alpha,\,\rm{obs}},
\label{eqLLyalpha}
\end{equation}
where $u_{\rm{Ly}\,\alpha}$ is the Ly $\alpha$ energy density. We assume that one ionised cloud is penetrated by the electron beam. The electrons interact with the emission line photons. A cascade evolves, reprocesses the injected energy into gamma rays, that escape in the direction of the incident electron beam, and forms a distinct emerging spectrum. As will be seen in the subsequent section, this cascaded spectrum, on top of the broadband SSC emission, can provide an explanation for the gamma-ray spectrum of Mrk\,501 from MJD 56857.98, as measured with the MAGIC telescopes.\\

\subsection{Very high energy gamma rays resulting from an electron-beam-driven pair cascade}\label{SubsecCascade}

\begin{floatingfigure}[l]{0.3\textwidth}
\centering
\includegraphics[width=0.3\textwidth]{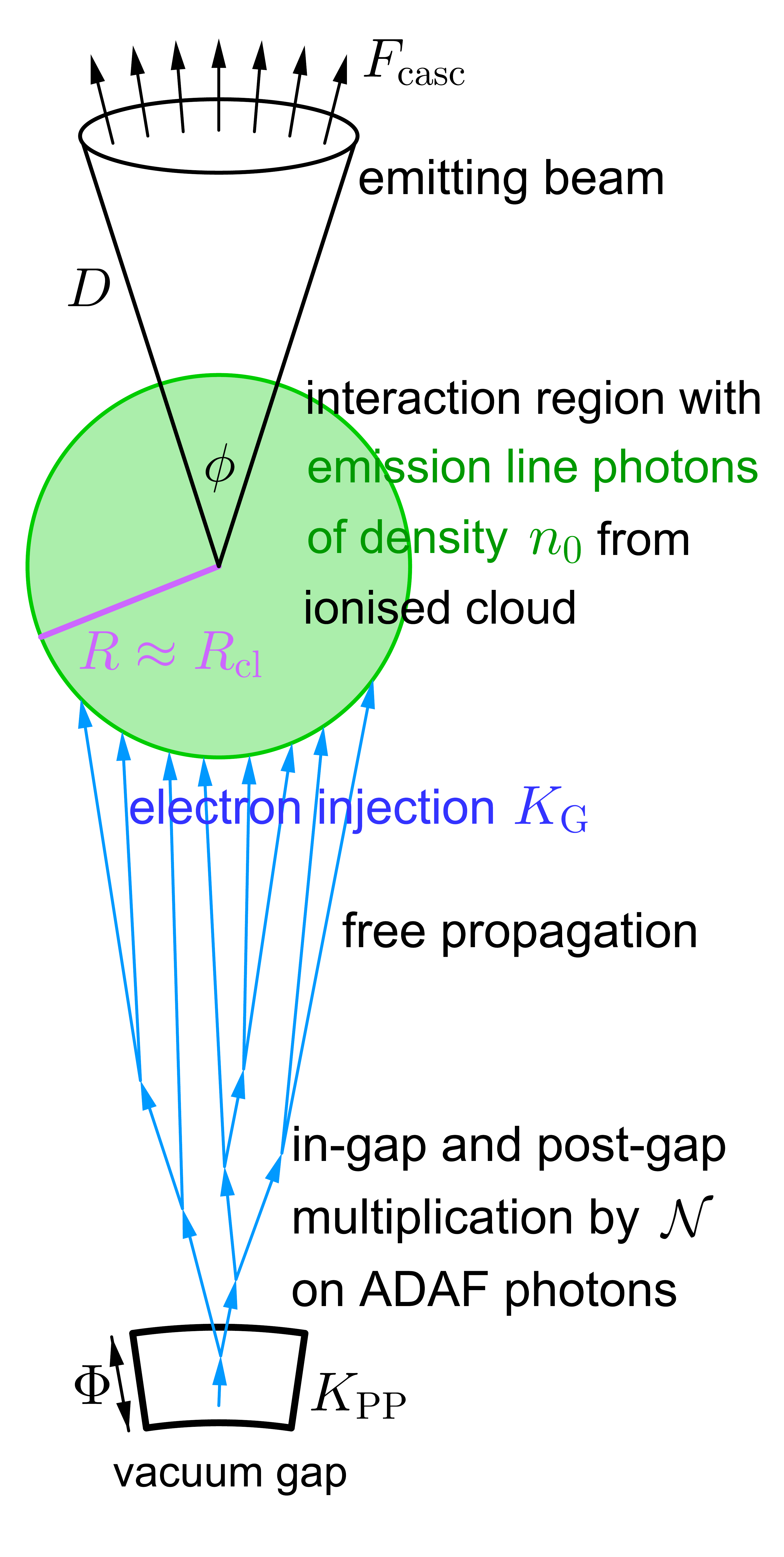}
\caption{Sketch of the scenario considered for the electron beam emission line photon interaction in Mrk\,501.}
\label{FigSketch}
\end{floatingfigure}
We briefly outline our approach to determine the emerging cascaded spectrum. We consider a spherical interaction region with radius $R$, in which we have homogeneous, isotropic and time-independent electron, gamma-ray photon and soft photon densities, $N(\gamma)$, $n_\gamma(x_\gamma)$ and $n_0(x)$, respectively. Into this region, electrons are injected with the spectral rate $\dot N_{\rm i}(\gamma)$. Electrons escape on the timescale $R/c$ and interact by IC scattering on the line photons. By this, the electrons lose energy, but they remain in the system. The IC scattering creates gamma-ray photons, some of which escape from the system and can be observed from Earth. The gamma-ray photons also collide with the soft photons and pair produce electrons and positrons. We first assume that the gamma-ray photon density is low enough for the cascade being linear, and later on (in Sect. \ref{Subsec}) this assumption is validated.

To describe the linear cascade, we specify all energy loss and gain rates of the electrons and of the gamma-ray photons, respectively \cite{1988ApJ...335..786Z,2021A&A...646A.115W}. After assuming steady state, we arrive at an equation $N(\gamma) = \mathcal{F}(n_0, \dot N_{\rm i}(\gamma), R, N(\gamma), \gamma)$. For given $n_0$, $\dot N_{\rm i}$ and $R$ and after choosing an initial $N$, it can be solved iteratively for the electron distribution $N$. Discretising $\gamma$ into a grid of 140 points, beginning at $\gamma_{{\mathrm{i}},\,0}$ and iterating towards lower energies, we achieve approximate convergence after a maximum number of $\approx 100$ iteration steps \cite{2021A&A...646A.115W}. At TeV energies, $N$ is shaped mainly due to the injected electrons. At GeV energies, escape dominates. From the electron density, we obtain the gamma-ray photon density \cite{2021A&A...646A.115W}, which has a peak at $x_\gamma \approx \gamma_{\mathrm{mean}}$ due to IC up-scattering of the soft photons by the injected electrons. Below this peak, the cascaded spectrum shows absorption troughs due to the maxima of the pair production cross section contributions of the various lines. Using $\phi$ for the opening angle of the electron and gamma-ray beam, and the luminosity distance $D = 149.4 \, \rm{Mpc}$, the observed cascaded flux density $F_{\rm{casc}}$ is determined \cite{2021A&A...646A.115W}.

As a first step, the quasi-simultaneous broadband multi-wavelength SED of Mrk\,501 on MJD 56857.98 is modelled with an SSC approach \cite{2020A&A...637A..86M}. As a second step, as shown in Fig. \ref{FigMrk501SED}, the cascade emission is added to the SSC contribution in order to properly describe the narrow spectral component centred at $\sim 3\,\rm TeV$ by adjusting the parameters $\phi$, $R$, $u_{\rm{lines,\,tot}}$, $K_{\rm{G}}$, $\varsigma$ and $\gamma_{\mathrm{mean}}$. After a good fit is achieved with $\phi = 1.8\,^{\circ}$, $R = 3.0 \cdot 10^{11} \, \rm{m}$, $u_{\rm{lines,\,tot}} = 6.5 \, \rm{J}/\rm{m}^3$, $K_{\rm{G}} = 3.3 \cdot 10^4 \, \rm{s^{-1} m^{-3}}$, $\gamma_{\mathrm{mean}} = 3.4 \cdot 10^{12} \, {\rm{eV}} / (m_{\rm{e}} c^2)$ and $\varsigma = 0.23 \, \gamma_{\mathrm{mean}}$, the fitting parameters are verified to comply with the setting described in Sect. \ref{SubsecSetting}.\\

\subsection{Consistency check}\label{Subsec}

From the obtained $u_{\rm{lines,\,tot}}$, and knowing about the various line contributions, we extract $u_{\rm{Ly}\,\alpha}$ \cite{2021A&A...646A.115W}. With this and the observed Ly $\alpha$ luminosity, we solve Eq. \ref{eqLLyalpha} for the cloud radius and obtain $R_{\rm cl} \approx 1.8 \cdot 10^{11} \, {\rm m}$, that is comparable within one order of magnitude with the obtained $R$. Having determined $u_{\rm{lines,\,tot}}$ and $R_{\rm cl}$ and using Eq. \ref{eqLlinestot2}, we obtain the luminosity of all emission line clouds $L_{\rm{lines,\,tot}} \approx 7.8 \cdot 10^{33} \, \rm{W}$.

According to Eq. \ref{eqLlinestot1}, this luminosity is supplied by the accretion flow. For $\xi = 0.1$ (canonical irradiation) and $\xi = 0.01$ (weak irradiation), we determine pairs of $\dot m$ and $T_{\rm e}$, such that Eq. \ref{eqLlinestot1} is satisfied \cite{2021A&A...646A.115W}. For $\dot m \in [1.0 \cdot 10^{-4}, 8.0 \cdot 10^{-4}]$, we find $T_{\rm e}/\rm{K} \in [6.4 \cdot 10^{9}, 1.3 \cdot 10^{10}]$ in the $\xi = 0.1$ case and $T_{\rm e}/\rm{K} \in [8.4 \cdot 10^{9}, 1.7 \cdot 10^{10}]$ in the $\xi = 0.01$ case, substantiating that the flow is in the ADAF regime.

For both $\xi$ cases, we use the pairs of $\dot m$ and $T_{\rm e}$ to determine the density of pair-production-capable ADAF photons, and obtain $n_{{\rm PP}}/{\rm m}^{-3} \in [2.8 \cdot 10^9, 4.7 \cdot 10^9]$ in the $\xi = 0.01$ case \cite{2021A&A...646A.115W}. For $\xi = 0.1$, the lower $T_{\rm e}$ results in $n_{{\rm PP}}$ being about a factor 0.01 lower. With Eq. \ref{eqPP}, we get the materialisation rate $K_{\rm{PP}}/\rm{s^{-1} m^{-3}} \in [0.031, 0.088]$ for $\xi = 0.01$, and values a factor $10^{-4}$ smaller in the $\xi = 0.1$ case. With $K_{\rm{G}}$ from the cascade modelling and with $K_{\rm{PP}}$, we use Eq. \ref{eqK} to obtain $\mathcal{N} \approx 10^{6}$ for $\xi = 0.01$, which is in between of magnetospheric model estimates \cite{Broderick,LevinsonCerutti}, and $\mathcal{N} \approx 10^{10}$ in the $\xi = 0.1$ case. $\mathcal{N}$, $\varsigma$ and $\gamma_{\mathrm{mean}}$ determine the necessary gap potential drop through Eq. \ref{eqPhi}, yielding a reasonable $\Phi = 5.7 \cdot 10^{18} \, \rm{V}$ \cite[cf.][]{2011ApJ...730..123L,HirotaniPu1} in the $\xi = 0.01$ case and an extreme $5.7 \cdot 10^{22} \, \rm{V}$ for the canonical $\xi$. Our model thus argues for a low reprocessing fraction of $\xi = 0.01$.

From $K_{\rm{G}}$, $\varsigma$ and $\gamma_{\mathrm{mean}}$ we determine the power $P_{\rm i} = 1.9 \cdot 10^{33} \, {\rm W}$ that is injected into the cascade with the line photons, and originating from the output of the vacuum gap. This is about $0.1 \, \%$ of the Blandford-Znajek power estimate, and confirms the simulated power output of a 10\,h-lasting vacuum gap discharge \cite{LevinsonCerutti}.

\section{Cascade on broad line region photons in 3C 279}\label{Sec3C279}

A conceptually similar approach is pursued in the case of 3C\,279. The location of its high-energy (HE) gamma-ray emission region can only be determined with indirect arguments based on variability, multi-wavelength SED modelling, constraints on the pair absorption optical depth, and the comparison of flare decay with cooling times \cite[e.g.][]{2019ApJ...877...39M,2021MNRAS.500.5297A}. Studies based on these arguments did, however, yield no consensus about the number and locations of gamma-ray emission regions in FSRQs. Instead, the emission regions seem to be multiple and non-stationary \cite[e.g.][]{2020NatCo..11.4176S,2021MNRAS.500.5297A}.
\begin{floatingfigure}[l]{0.55\textwidth}
\centering
\includegraphics[width=0.55\textwidth]{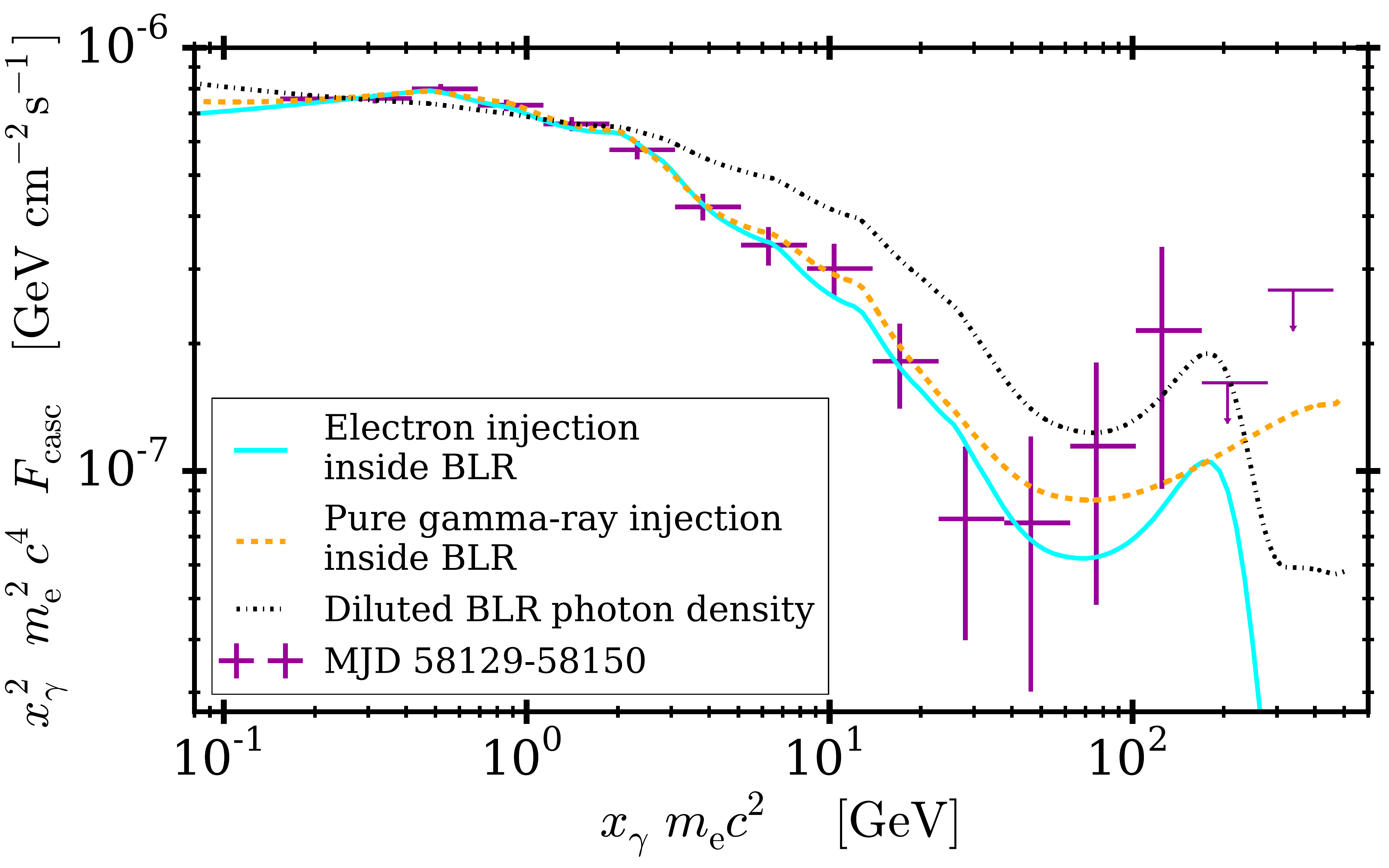}
\caption{Energy flux density versus photon energy of the {\it Fermi}-LAT observation (violet markers) of 3C\,279, as well as for different theoretical model cases. Models with injection and cascading in the BLR photon field are depicted by the cyan solid and by the orange dashed line, while injection outside of the BLR is shown by the black dash-dotted line.}
\label{Fig3C279SED}
\end{floatingfigure}
In contrast to studies considering pure pair absorption of intrinsic spectra, we compute the full cascade with inclusion of escape and IC up-scattering of emission line photons by the injected and pair-produced electrons \cite{W.S.andM.submittedtoApJ}. In three cases, we inject ultra-relativistic electrons and HE photons with hard spectral shape into a broad line region (BLR) photon field, consisting of UV and soft X-ray lines \cite{1997A&AS..125..149D,2012ApJ...744...99L,2017MNRAS.464..152A}. Similarly to the approach described in Sect. \ref{SubsecCascade}, we determine the electron energy distribution and the escaping gamma-ray flux density. By choosing the emission line strengths within reasonable borders, we can fit the modelled HE SED to the {\it Fermi} Large Area Telescope (LAT) spectrum obtained during a flaring period from MJD 58129 to MJD 58150 (January 2018). As seen in the cyan solid and orange dashed line in Fig. \ref{Fig3C279SED}, the result is quite independent of the exact composition of the injected species. Depending on the injection, we achieve reduced chi squares $\chi^2_{\rm red} = 0.90, 0.62$, and $0.83$ with the numbers of degrees of freedom (NDF) being 7, 9, and 7, respectively. A logparabolic fit yields a minimum $\chi^2_{\rm red} = 1.2$ with NDF = 11 \cite{W.S.andM.submittedtoApJ}.

We also model injection and cascading outside of the BLR by employing a diluted emission line photon density, but the same escape radius. By this, pair absorption is reduced and escape has a stronger influence. With these prerequisites and with non-extreme spectral injection rates, we find it impossible to achieve a good fit, as the SED trough above 10\,GeV cannot be met, without emission lines strong enough, cf. black dash-dotted line in Fig \ref{Fig3C279SED}. Therefore, we conclude that the hints to the fine structure in the LAT SED are a signature of the interplay between pair absorption and radiation reprocessing from the edge of the BLR.

\section{Summary}\label{SecSummary}

High-precision gamma-ray observations reveal SED substructures beyond the predictions of spherical blob models, but in line with the predictions of IC pair cascade models in external radiation fields. To show this, we have developed a robust algorithm to determine the electron and photon distributions of steady-state, linear IC pair cascades with escape. 
In the case of Mrk\,501, an electron beam from a sporadically active vacuum gap interacting with emission line photons from an ionised cloud can explain a narrow feature at TeV energies observed during a flare in July 2014, if the field of clouds reprocesses $1\,\%$ of the illuminating ADAF luminosity. The hints of a trough in the {\it Fermi}-LAT SED of 3C\,279 from January 2018 can be explained by a cascade taking place in or at the edge of the BLR.

\acknowledgments
C. W. is thankful for support of the Universität zu K\"oln and of the German Bundesministerium f\"{u}r Arbeit und Soziales. We also thank the developers of numpy \cite{10.5555/2886196}, matplotlib \cite{2007CSE.....9...90H} and the collaborative Chianti project by the George Mason University, the University of Michigan, the University of Cambridge and the NASA Goddard Space Flight Center.

\bibliographystyle{JHEP}
\bibliography{Proceeding-2021-06-30.bib}

\end{document}